\catcode`\@=11
\expandafter\ifx\csname @iasmacros\endcsname\relax
	\global\let\@iasmacros=\par
\else	\immediate\write16{}
	\immediate\write16{Warning:}
	\immediate\write16{You have tried to input iasmacrosy more than once.}
	\immediate\write16{}
	\endinput
\fi
\catcode`\@=12


\def\rmb{\seventeenrm}

\def\singlespace{\baselineskip=\normalbaselineskip}
\def\halfspace{\baselineskip=1.5\normalbaselineskip}
\def\doublespace{\baselineskip=2\normalbaselineskip}


\def\AB{\bigskip\parindent=40pt
        \centerline{\bf ABSTRACT}\medskip\halfspace\narrower}
\def\AE{\bigskip\nonarrower\doublespace}
\def\nonarrower{\advance\leftskip by-\parindent
	\advance\rightskip by-\parindent}


\def\boxit#1{\vbox{\hrule\hbox{\vrule\kern3pt
	\vbox{\kern3pt#1\kern3pt}\kern3pt\vrule}\hrule}}

\def\hence{\leavevmode\hbox{\bf .\raise5.5pt\hbox{.}.} }

\def\dalemb#1#2{{\vbox{\hrule height.#2pt
	\hbox{\vrule width.#2pt height#1pt \kern#1pt \vrule width.#2pt}
	\hrule height.#2pt}}}
\def\gtorder{\mathrel{\raise.3ex\hbox{$>$}\mkern-14mu
             \lower0.6ex\hbox{$\sim$}}}
\def\ltorder{\mathrel{\raise.3ex\hbox{$<$}\mkern-14mu
             \lower0.6ex\hbox{$\sim$}}}

\newdimen\fullhsize
\newbox\leftcolumn
\def\twoup{\hoffset=-.5in \voffset=-.25in
  \hsize=4.75in \fullhsize=10in \vsize=6.9in
  \def\fullline{\hbox to\fullhsize}
  \let\lr=L
  \output={\if L\lr
        \global\setbox\leftcolumn=\columnbox\global\let\lr=R \advancepageno
      \else \doubleformat \global\let\lr=L\fi
    \ifnum\outputpenalty>-20000 \else\dosupereject\fi}
  \def\doubleformat{\shipout\vbox{
    \fullline{\box\leftcolumn\hfil\columnbox}\advancepageno}}
  \def\columnbox{\leftline{\vbox{\makeheadline\pagebody\makefootline}}}
  \tolerance=1000 }

\catcode`\@=11					



\font\fiverm=cmr5				
\font\fivemi=cmmi5				
\font\fivesy=cmsy5				
\font\fivebf=cmbx5				

\skewchar\fivemi='177
\skewchar\fivesy='60


\font\sixrm=cmr6				
\font\sixi=cmmi6				
\font\sixsy=cmsy6				
\font\sixbf=cmbx6				

\skewchar\sixi='177
\skewchar\sixsy='60


\font\sevenrm=cmr7				
\font\seveni=cmmi7				
\font\sevensy=cmsy7				
\font\sevenit=cmti7				
\font\sevenbf=cmbx7				

\skewchar\seveni='177
\skewchar\sevensy='60


\font\eightrm=cmr8				
\font\eighti=cmmi8				
\font\eightsy=cmsy8				
\font\eightit=cmti8				
\font\eightbf=cmbx8				

\skewchar\eighti='177
\skewchar\eightsy='60


\font\ninei=cmmi9
\font\ninesy=cmsy9

\skewchar\ninei='177
\skewchar\ninesy='60


\font\tenrm=cmr10				
\font\teni=cmmi10				
\font\tensy=cmsy10				
\font\tenex=cmex10				
\font\tenit=cmti10				
\font\tensl=cmsl10				
\font\tenbf=cmbx10				
\font\tentt=cmtt10				
\font\tenss=cmss10				
\font\tensc=cmcsc10				
\font\tenbi=cmmib10				

\skewchar\teni='177
\skewchar\tenbi='177
\skewchar\tensy='60

\def\tenpoint{\ifmmode\err@badsizechange\else
	\textfont0=\tenrm \scriptfont0=\sevenrm \scriptscriptfont0=\fiverm
	\textfont1=\teni  \scriptfont1=\seveni  \scriptscriptfont1=\fivemi
	\textfont2=\tensy \scriptfont2=\sevensy \scriptscriptfont2=\fivesy
	\textfont3=\tenex \scriptfont3=\tenex   \scriptscriptfont3=\tenex
	\textfont4=\tenit \scriptfont4=\sevenit \scriptscriptfont4=\sevenit
	\textfont5=\tensl
	\textfont6=\tenbf \scriptfont6=\sevenbf \scriptscriptfont6=\fivebf
	\textfont7=\tentt
	\textfont8=\tenbi \scriptfont8=\seveni  \scriptscriptfont8=\fivemi
	\def\rm{\tenrm\fam=0 }%
	\def\it{\tenit\fam=4 }%
	\def\sl{\tensl\fam=5 }%
	\def\bf{\tenbf\fam=6 }%
	\def\tt{\tentt\fam=7 }%
	\def\ss{\tenss}%
	\def\sc{\tensc}%
	\def\bmit{\fam=8 }%
	\rm\setparameters\setbaselines\fi}


\font\twelverm=cmr12				
\font\twelvei=cmmi12				
\font\twelvesy=cmsy10	scaled\magstep1		
\font\twelveex=cmex10	scaled\magstep1		
\font\twelveit=cmti12				
\font\twelvesl=cmsl12				
\font\twelvebf=cmbx12				
\font\twelvett=cmtt12				
\font\twelvess=cmss12				
\font\twelvesc=cmcsc10	scaled\magstep1		
\font\twelvebi=cmmib10	scaled\magstep1		

\skewchar\twelvei='177
\skewchar\twelvebi='177
\skewchar\twelvesy='60

\def\twelvepoint{\ifmmode\err@badsizechange\else
	\textfont0=\twelverm \scriptfont0=\eightrm \scriptscriptfont0=\sixrm
	\textfont1=\twelvei  \scriptfont1=\eighti  \scriptscriptfont1=\sixi
	\textfont2=\twelvesy \scriptfont2=\eightsy \scriptscriptfont2=\sixsy
	\textfont3=\twelveex \scriptfont3=\tenex   \scriptscriptfont3=\tenex
	\textfont4=\twelveit \scriptfont4=\eightit \scriptscriptfont4=\sevenit
	\textfont5=\twelvesl
	\textfont6=\twelvebf \scriptfont6=\eightbf \scriptscriptfont6=\sixbf
	\textfont7=\twelvett
	\textfont8=\twelvebi \scriptfont8=\eighti  \scriptscriptfont8=\sixi
	\def\rm{\twelverm\fam=0 }%
	\def\it{\twelveit\fam=4 }%
	\def\sl{\twelvesl\fam=5 }%
	\def\bf{\twelvebf\fam=6 }%
	\def\tt{\twelvett\fam=7 }%
	\def\ss{\twelvess}%
	\def\sc{\twelvesc}%
	\def\bmit{\fam=8 }%
	\rm\setparameters\setbaselines\fi}


\font\fourteenrm=cmr12	scaled\magstep1		
\font\fourteeni=cmmi12	scaled\magstep1		
\font\fourteensy=cmsy10	scaled\magstep2		
\font\fourteenex=cmex10	scaled\magstep2		
\font\fourteenit=cmti12	scaled\magstep1		
\font\fourteensl=cmsl12	scaled\magstep1		
\font\fourteenbf=cmbx12	scaled\magstep1		
\font\fourteentt=cmtt12	scaled\magstep1		
\font\fourteenss=cmss12	scaled\magstep1		
\font\fourteensc=cmcsc10 scaled\magstep2	
\font\fourteenbi=cmmib10 scaled\magstep2	

\skewchar\fourteeni='177
\skewchar\fourteenbi='177
\skewchar\fourteensy='60

\def\fourteenpoint{\ifmmode\err@badsizechange\else
	\textfont0=\fourteenrm \scriptfont0=\tenrm \scriptscriptfont0=\sevenrm
	\textfont1=\fourteeni  \scriptfont1=\teni  \scriptscriptfont1=\seveni
	\textfont2=\fourteensy \scriptfont2=\tensy \scriptscriptfont2=\sevensy
	\textfont3=\fourteenex \scriptfont3=\tenex \scriptscriptfont3=\tenex
	\textfont4=\fourteenit \scriptfont4=\tenit \scriptscriptfont4=\sevenit
	\textfont5=\fourteensl
	\textfont6=\fourteenbf \scriptfont6=\tenbf \scriptscriptfont6=\sevenbf
	\textfont7=\fourteentt
	\textfont8=\fourteenbi \scriptfont8=\tenbi \scriptscriptfont8=\seveni
	\def\rm{\fourteenrm\fam=0 }%
	\def\it{\fourteenit\fam=4 }%
	\def\sl{\fourteensl\fam=5 }%
	\def\bf{\fourteenbf\fam=6 }%
	\def\tt{\fourteentt\fam=7}%
	\def\ss{\fourteenss}%
	\def\sc{\fourteensc}%
	\def\bmit{\fam=8 }%
	\rm\setparameters\setbaselines\fi}


\font\seventeenrm=cmr10 scaled\magstep3		


\newdimen\rp@
\newcount\@basestretchnum
\newskip\@baseskip
\newskip\headskip
\newskip\footskip


\def\setparameters{\rp@=.1em
	\headskip=24\rp@
	\footskip=\headskip
	\delimitershortfall=5\rp@
	\nulldelimiterspace=1.2\rp@
	\scriptspace=0.5\rp@
	\abovedisplayskip=10\rp@ plus3\rp@ minus5\rp@
	\belowdisplayskip=10\rp@ plus3\rp@ minus5\rp@
	\abovedisplayshortskip=5\rp@ plus2\rp@ minus4\rp@
	\belowdisplayshortskip=10\rp@ plus3\rp@ minus5\rp@
	\normallineskip=\rp@
	\lineskip=\normallineskip
	\normallineskiplimit=0pt
	\lineskiplimit=\normallineskiplimit
	\jot=3\rp@
	\setbox0=\hbox{\the\textfont3 B}\p@renwd=\wd0
	\skip\footins=12\rp@ plus3\rp@ minus3\rp@
	\skip\topins=0pt plus0pt minus0pt}


\def\setbaselines{\maxdepth=4\rp@\baselinestretch=\@basestretchnum}


\def\baselinestretch{\afterassignment\@basestretch\@basestretchnum}
\def\@basestretch{%
	\@baseskip=12\rp@ \divide\@baseskip by1000
	\normalbaselineskip=\@basestretchnum\@baseskip
	\baselineskip=\normalbaselineskip
	\bigskipamount=\the\baselineskip
		plus.25\baselineskip minus.25\baselineskip
	\medskipamount=.5\baselineskip
		plus.125\baselineskip minus.125\baselineskip
	\smallskipamount=.25\baselineskip
		plus.0625\baselineskip minus.0625\baselineskip
	\setbox\strutbox=\hbox{\vrule height.708\baselineskip
		depth.292\baselineskip width0pt }}



\def\makeheadline{\vbox to0pt{\baselinestretch=1000
	\vskip-\headskip \vskip1.5pt
	\line{\vbox to\ht\strutbox{}\the\headline}\vss}\nointerlineskip}

\def\makefootline{\baselineskip=\footskip\line{\the\footline}}

\def\big#1{{\hbox{$\left#1\vbox to8.5\rp@ {}\right.\n@space$}}}
\def\Big#1{{\hbox{$\left#1\vbox to11.5\rp@ {}\right.\n@space$}}}
\def\bigg#1{{\hbox{$\left#1\vbox to14.5\rp@ {}\right.\n@space$}}}
\def\Bigg#1{{\hbox{$\left#1\vbox to17.5\rp@ {}\right.\n@space$}}}


\mathchardef\alpha="710B
\mathchardef\beta="710C
\mathchardef\gamma="710D
\mathchardef\delta="710E
\mathchardef\epsilon="710F
\mathchardef\zeta="7110
\mathchardef\eta="7111
\mathchardef\theta="7112
\mathchardef\iota="7113
\mathchardef\kappa="7114
\mathchardef\lambda="7115
\mathchardef\mu="7116
\mathchardef\nu="7117
\mathchardef\xi="7118
\mathchardef\pi="7119
\mathchardef\rho="711A
\mathchardef\sigma="711B
\mathchardef\tau="711C
\mathchardef\upsilon="711D
\mathchardef\phi="711E
\mathchardef\chi="711F
\mathchardef\psi="7120
\mathchardef\omega="7121
\mathchardef\varepsilon="7122
\mathchardef\vartheta="7123
\mathchardef\varpi="7124
\mathchardef\varrho="7125
\mathchardef\varsigma="7126
\mathchardef\varphi="7127
\mathchardef\imath="717B
\mathchardef\jmath="717C
\mathchardef\ell="7160
\mathchardef\wp="717D
\mathchardef\partial="7140
\mathchardef\flat="715B
\mathchardef\natural="715C
\mathchardef\sharp="715D


\def\err@badsizechange{%
	\immediate\write16{--> Size change not allowed in math mode, ignored}}

\baselinestretch=1000
\tenpoint

\catcode`\@=12					

\twelvepoint
\doublespace
\overfullrule=0pt
{\nopagenumbers{
\rightline{IASSNS-HEP-99/01}
\rightline{~~~January, 1999}
\bigskip\bigskip
\centerline{\rmb   Higgs Mass Bounds in the Three- and } 
\centerline{\rmb  Six-Higgs Doublet Models for Family Structure}
\medskip
\centerline{\it Stephen L. Adler
}
\centerline{\bf Institute for Advanced Study}
\centerline{\bf Princeton, NJ 08540}
\bigskip\bigskip
\leftline{\it Send correspondence to:}
\medskip
{\singlespace\leftline{Stephen L. Adler}
\leftline{Institute for Advanced Study}
\leftline{Olden Lane, Princeton, NJ 08540}
\leftline{Phone 609-734-8051; FAX 609-924-8399; email adler@ias.edu}}
\bigskip\bigskip
}}
\vfill\eject
\pageno=2
\AB
We reanalyze our recently proposed mass matrix model   
based on spontaneously broken discrete chiral family symmetry, taking 
into account the additional flavor changing neutral current constraint 
implied by the bound on the $D_1-D_2$ mass difference, and including several   
corrections to our earlier analysis.  When combined, the $K_1-K_2$ and 
$D_1-D_2$ constraints 
force the masses of the Higgs particles that contribute most strongly 
to flavor changing neutral currents (the $\phi$ Higgs states) 
to lie above 17 TeV, well 
beyond the limit of validity of conventional perturbative Higgs physics.     
The analogous constraints on the masses of the $\eta$ Higgs states 
and the neutral   
pseudo Goldstone Higgs state depend on the mechanism for realizing 
small first family masses. If the $\eta$ Higgs 
is the primary contributor to second family masses, the pseudo Goldstone  
and $\eta$ Higgs states must have masses above 220 GeV, with numerical 
fits suggesting masses above 1 TeV, while if the $\eta$ 
Higgs is responsible solely for first family masses, the corresponding  
mass bounds drop to the range detectable at the LHC.   
We show that naturalness of small first family masses
favors the latter alternative, and give an illustrative mass matrix      
texture model.  
  
\AE
\bigskip\bigskip
\vfill\eject
\pageno=3
\centerline{{\bf I.~~Introduction}}
\bigskip
In a recent paper [1] we constructed extensions of the standard model, 
based on the hypothesis that the Higgs bosons also exhibit a threefold 
family structure, and that the flavor weak eigenstates are distinguished 
by a discrete $Z_6$ chiral symmetry that is spontaneously broken by the 
Higgs sector.  Two models were analyzed in [1], the first with one 
three-family set of Higgs doublets, and the second with two three-family 
sets of Higgs doublets.  In the three-Higgs doublet model, the leading 
cyclically symmetric approximation to the quark and lepton mass matrices 
has the ``democratic'' form with all matrix elements equal, leading to 
one massive and two massless fermion 
families.  In the six-Higgs doublet model, for a wide 
range of Higgs potential parameters, $CP$ is spontaneously broken, and this 
breaking simultaneously modifies the democratic Ansatz to give nonzero 
masses to an additional family (assumed in [1] to be the second family)   
in leading cyclic approximation.  Corrections to the cyclic approximation 
were used in [1] to give first family masses, and a nontrivial CKM matrix.  

In performing numerical fits to the data using the models of [1], we 
took into account bounds on flavor changing neutral currents solely 
through the constraint provided by the $K_1-K_2$ mass difference, 
which led to strong asymmetries in the fits 
between the up and down quark sectors.  
M. Peskin [2] has pointed out the importance 
of including in the analysis experimental bounds on the $D_1-D_2$ 
mass difference, which is the up quark sector analog of the $K_1-K_2$ 
mass difference constraint.  The purpose of this paper is 
to give the formulas and numerical results needed for this extension of 
the analysis of [1].  We also consider an alternative version of the 
model of [1], in which cyclic asymmetries in the $\phi$ Higgs couplings 
are responsible for second family masses, and the $\eta$ Higgs contributes  
significantly only to 
first family masses.  We show that this alternative is favored by requiring 
naturalness of small first family masses.  
In addition, we make the following three corrections 
to the model as originally formulated: (i) we correct the  
form of the CKM matrix, as pointed out in an Erratum [3] to [1], (ii) we 
include rephasings needed to make the diagonalized quark mass matrices 
positive real, and (iii) we correct combinatoric factors in the 
flavor changing neutral current amplitude (amounting to an overall factor 
of 2), and give a more accurate treatment of the hadronic 
matrix elements appearing in the flavor changing neutral current constraints. 

This paper is organized as follows.  In Sec. II we give a synopsis of results 
needed from [1], including the corrections (i) and (ii) noted above.  In 
Secs. III  and IV we analyze  $K_1-K_2$ and $D_1-D_2$ mixing induced by 
Higgs exchange, including the corrections (iii).   In Sec. III 
we give formulas for 
calculating the $\phi$ Higgs, the $\eta$ Higgs, and 
the pseudo Goldstone Higgs 
contributions, in the six-Higgs doublet model, to both the $K_1-K_2$ 
and $D_1-D_2$ mass differences.   In Sec. IV we use the formulas of 
Secs. II and III to derive a series of 
bounds on the Higgs masses, which are evaluated numerically    
using lattice and model calculations of the relevant hadronic  
matrix elements, for two possible mechanisms for realizing the first   
and second family masses.  Irrespective of this choice of mechanism, we 
find that the $\phi$ Higgs masses must be greater than 17 TeV, 
in accord with analyses [4] of generic multi-Higgs models.  This bound 
also extends, by specialization to the case in which the $\eta$ Higgs  
couplings vanish, to the CP conserving case of the three-Higgs
doublet model. 
In Sec. V we analyze the implications of requiring that small first 
family masses arise naturally, as opposed to arising by detailed 
cancellations between physically unrelated quantities, and give a   
simple mass matrix texture model corresponding to the case in which 
Yukawa coupling asymmetries are responsible for second family masses.  
In Sec. VI we 
repeat the numerical fits of [1], taking into account the results 
derived in the preceding sections, and summarize our conclusions.   

\bigskip
\centerline{\bf II.~~Synopsis of Needed Results from the Six-Higgs Model}

The six-Higgs doublet model of [1] is based on 
the assumption that there are 
two discrete chiral families of Higgs bosons, $\phi_n$ and $\eta_n$, 
$n=1,2,3$.    
These are coupled to discrete chiral families of fermions 
to give a Lagrangian that is exactly discrete chiral invariant, 
and that is approximately invariant under cyclic permutation of the discrete 
chiral components.  The model is constructed so that the 
Higgs fields develop nonvanishing vacuum expectations in a CP violating 
phase, and it is assumed that the $\phi$ Higgs bosons  couple much more 
strongly to fermions than the $\eta$ Higgs bosons, and similarly   
for their corresponding expectations (denoted respectively by $\Omega_{\phi}$  
and $\Omega_{\eta}$).  
As a zeroth order approximation to the model, only the 
$\phi$ Higgs expectations are retained and cyclic permutation symmetry 
is assumed, leading to a ``democratic'' mass matrix with one massive and 
two massless families, and a CKM matrix of unity.  

Deviations from cyclic symmetry, and the $\eta$ Higgs expectations, 
are then added back as a perturbation, giving as the Lagrangian mass term
$${\cal L}_{\rm mass}=\sum_{f=u,d,e}{\overline f}_L^{\prime} 
g_{\phi}^f \Omega_{\phi}(3M^{(3)}+\sigma^f)f_R^{\prime}~~~.\eqno(1a)$$
Here  $M^{(3)}={\rm diag}(0,0,1)$ is the projector on the third family,   
$g_{\phi}^f$ is the $\phi$ Higgs Yukawa coupling for flavor $f$, 
and 
$\sigma^f$ is a $3 \times 3$ matrix of perturbations, given explicitly by 
$$\eqalign{
\sigma_{11}^f=&{1\over 3} \mu_{11}^f+\delta_3^f+\overline \omega \delta_2^f 
+\omega \delta_1^f~~~,\cr
\sigma_{22}^f=&{1 \over 3} \mu_{22}^f + 3R^f+\delta_3^f+\omega \delta_2^f
+\overline \omega \delta_1^f~~~,\cr
\sigma_{33}^f=&0~~~,\cr
\sigma_{\ell m}^f=&{1\over 3}\mu_{\ell m}^f,~~ \ell \neq m
~;~~~\omega\equiv \exp(2\pi i/3)~,~~~\overline \omega\equiv \omega^*~~~.\cr
}\eqno(1b)$$
In Eq.~(1b) the terms $\mu_{\ell m}$ arise from small deviations 
from cyclic symmetry 
in the Yukawa couplings of the $\phi$ Higgs bosons, the terms $\delta_n^f$ 
arise from deviations from cyclic symmetry in 
the $\phi$ Higgs expectations, and the term $R^f$ arises from   
contributions to the mass matrix of the weakly coupled 
$\eta$ Higgs expectations.  Further details of the structure of $\sigma^f$ 
are given in Eqs.~(38b) through (39c) of [1] and are used in Secs. V and  
VI below.  However, the only property needed for the analytic 
calculations of Sec. III is that, since the Yukawa coupling asymmetries 
are all real because the Lagrangian in the 
six-Higgs doublet model is assumed to be CP invariant, the first 
order perturbations $\sigma^f$ obey 
$$\sigma_{12}^f=\sigma_{21}^{f*} ~~~.\eqno(1c)$$ 
This restriction holds even though the model of [1] chooses a CP violating 
ground state.  

Defining
$$M_f^{\prime} = 3M^{(3)}+\sigma^f~~~,\eqno(2a)$$
we then [1] construct the bi-unitary transformation matrices 
$U_L^f~,~~U_R^f$ for which $U_L^f M_f^{\prime} U_R^{f \dagger}$ is 
diagonal, with the eigenvalues ordered in absolute value.  
The fermion mass eigenstate basis, up to rephasings to be discussed,  
is related to the primed basis 
by 
$$\eqalign{
f_L^{\prime}=&U_L^{f\dagger}f_L^{\rm mass}~~~,\cr
f_R^{\prime}=&U_R^{f\dagger}f_R^{\rm mass}~~~,~~f=u,d,e~~~,
\cr} \eqno(2b)$$
and the CKM matrix $U_{\rm CKM}$ is given [3] by 
$$U_{\rm CKM}=U_L^uU_L^{d\dagger}~~~.\eqno(2c)$$

Since Eq.~(2a) defines a degenerate perturbation problem, the matrices 
$U_{L,R}^f$ are constructed [1] in two stages: first  the $2\times 2$ 
submatrix of $M_f^{\prime}$ spanned by the first two families is 
diagonalized exactly, and then the solution to this problem is used to 
perturbatively construct the full $3\times 3$ diagonalizing matrices.  
Because the analysis of flavor changing neutral current effects in the  
next section ignores third family mixings, it suffices for this analysis to
discuss only the $2\times 2$ submatrix diagonalization problem.  
Suppressing for the time being the flavor superscript $f$, we define the 
$2 \times 2$ submatrix $m$ by 
$$m = \pmatrix{ \sigma_{11} & \sigma_{12} \cr
\sigma_{21} & \sigma_{22} \cr } ~~~,\eqno(3a)$$ 
which is brought to diagonal (but not necessarily real) 
form by matrices $V_{L,R}$, 
$$V_L m V_R^{\dagger}=\pmatrix{\kappa_1 & 0\cr
0& \kappa_2 \cr}~~~,\eqno(3b)$$
with $|\kappa_1| \leq |\kappa_2|$.  An explicit construction of $V_{L,R}$ 
is given in Appendix B of [1]; the results obtained there can be simplified 
by using the symmetry of Eq.~(1c) above, which (in terms of the quantities 
defined in Appendix B of [1]) implies that 
$$\eqalign{
A_L=&A_R=|\sigma_{11}|^2+|\sigma_{12}|^2 \equiv A~~~,\cr
B_L=&B_R=|\sigma_{22}|^2+|\sigma_{12}|^2 \equiv B~~~,\cr
z_L=&(\sigma_{11}^*+\sigma_{22})\sigma_{12}^*~,~~
z_R=(\sigma_{11}+\sigma_{22}^*)\sigma_{12}^*~~~,\cr
|z_L|=&|z_R|\equiv |z|~~~.\cr
}\eqno(4a)$$
These relations, together with the results in Appendix B of [1], imply 
that 
$$V_{L,R}=\pmatrix{\cos \Theta & -\exp(-i\phi_{L,R}) \sin \Theta \cr
\exp(i \phi_{L,R}) \sin \Theta & \cos \Theta \cr }~~~,\eqno(4b)$$
with
$$\exp(i\phi_{L,R})=z_{L,R}/|z|~~~,\eqno(4c)$$
and 
$$\Theta={1 \over 2}{\tan}^{-1}\left({-2|z|\over A-B} \right)~~~.\eqno(4d)$$

Although the construction just given suffices for the computation of the 
magnitudes of the CKM matrix elements, the calculation of the Higgs 
exchange amplitude in the next section 
requires care in the choices of phases.  When we rephase the physical 
mass eigenstates $f_{L,R}^{\rm mass}$, the matrices $V^f_{L,R}$ transform 
according to 
$$V^f_{L,R} \to {\hat V}^f_{L,R} = D^f_{L,R} V^f_{L,R}~~~,\eqno(5a)$$
with $D^f_{L,R}$ diagonal unitary matrices.  A correct choice of phases 
requires that the diagonalized mass matrix   
$${\hat V}_L^f m^f {\hat V}_R^{f \dagger}~~~\eqno(5b)$$
be real and positive; in other words, the matrices $D^f_{L,R}$ must be chosen 
to 
absorb the phases of the diagonal matrix elements $\kappa_{1,2}^f$ 
on the right of Eq.~(3b) above.  In addition, restricting ourselves now 
to the up and down quark flavor sectors, we shall require that the phase 
choices for the physical states put the matrix 
${\hat V}_L^u {\hat V}_L^{d \dagger}$, 
which is the $2 \times 2$ submatrix of the rephased 
CKM matrix when third family  
mixings are neglected, into the standard real form 
$$\pmatrix {c_{12} & s_{12} \cr -s_{12} & c_{12} \cr }~~~,\eqno(5c)$$
with $s_{12}$ and $c_{12}$ both nonnegative.  
These two phase requirements together fix the rephasing 
matrices $D_{L,R}^f$ up to 
an irrelevant overall phase.  

To carry this construction out explicitly, 
we write Eq.~(3b), in the up and down quark sectors, as 
$$V_L^f m^f V_R^{f \dagger} = \pmatrix { |\kappa_1^f| \exp(i\theta_1^f) 
& 0 \cr 0 & |\kappa_2^f| \exp(i\theta_2^f) \cr }~,~~f=u,d~~~,\eqno(6a)$$
and we write the adjoint of the 
$2 \times 2$ CKM matrix computed before rephasing as 
$$V_L^d V_L^{u \dagger} = \pmatrix {c_{12}\exp(i\theta_{11}) & 
-s_{12} \exp(i\theta_{12}) \cr s_{12}\exp(i\theta_{21}) & 
c_{12} \exp(i\theta_{22}) \cr}~~~,\eqno(6b)$$
with unitarity imposing the conditions 
$$\eqalign{
&c_{12}^2+s_{12}^2=1 ~~~,\cr
&\theta_{11}+\theta_{22}=\theta_{12}+\theta_{21}~~ ({\rm mod}~ 2\pi)~~~.\cr
}\eqno(6c)$$
Then a simple calculation gives 
$$\eqalign{
D_L^d=&{\rm diag}[\exp(-i\theta_{11}),\exp(-i\theta_{21})]~~~,\cr
D_R^d=&{\rm diag}[\exp(i\theta_1^d-i\theta_{11}),
\exp(i\theta_2^d-i\theta_{21})]~~~,\cr
D_L^u=&{\rm diag}[1,\exp(-i\theta_{21}+i\theta_{22})]~~~,\cr
D_R^u=&{\rm diag}[\exp(i \theta_1^u),
\exp(i\theta_2^u-i\theta_{21}+i\theta_{22})]~~~.\cr
}\eqno(6d)$$
Corresponding to these, we find from Eqs.~(4b) and (5a) that 
$${\hat V}_L^d=\pmatrix{\exp(-i\theta_{11})c_d  
&-\exp(-i\phi_L^d-i\theta_{11})s_d \cr              
\exp(i\phi_L^d-i\theta_{21})s_d 
& \exp(-i \theta_{21}) c_d \cr}~~~,\eqno(7a)$$
$${\hat V}_R^d=\pmatrix{\exp(-i\theta_{11}+i\theta_1^d)c_d  
&-\exp(-i\phi_R^d-i\theta_{11}+i\theta_1^d)s_d \cr              
\exp(i\phi_R^d-i\theta_{21}+i\theta_2^d)s_d 
& \exp(-i \theta_{21}+i\theta_2^d) c_d \cr}~~~,\eqno(7b)$$
$${\hat V}_L^u=\pmatrix{c_u  
&-\exp(-i\phi_L^u)s_u \cr              
\exp(i\phi_L^u-i\theta_{21}+i\theta_{22})s_u 
& \exp(-i \theta_{21}+i\theta_{22}) c_u \cr}~~~,\eqno(7c)$$
$${\hat V}_R^u=\pmatrix{\exp(i\theta_1^u)c_u  
&-\exp(-i\phi_R^u+i\theta_1^u)s_u \cr              
\exp(i\phi_R^u-i\theta_{21}+i\theta_{22}+i\theta_2^u)s_u 
& \exp(-i \theta_{21}+i\theta_{22}+i\theta_2^u) c_u \cr}~~~,\eqno(7d)$$
with $c_{d,u},~s_{d,u}$ defined in terms of the angle $\Theta$ of 
Eq.~(4d) by
$$c_{d,u}=\cos(\Theta_{d,u})~,~~~s_{d,u}=\sin(\Theta_{d,u})~~~.\eqno(7e) $$
Equations (4a-d), (6a,~b), and (7a-e) provide our starting point for 
calculating the Higgs exchange contributions to flavor changing 
neutral current processes.

\bigskip
\centerline{\bf III.~~Analysis of $K_1-K_2$ and $D_1-D_2$ Mixing } 
\centerline{\bf Induced by Higgs Exchange}

We begin by extending the formulas of [1] for the Higgs 
exchange contribution to the $K_1-K_2$ mass difference to the case 
when the $\eta$ Higgs and pseudo Goldstone Higgs contributions 
are also included, using the rephased matrices ${\hat V}_{L,R}^{u,d}$ of   
Eqs.~(7a-e).  Our starting point is Eq.~(45c) of [1] for the $\Delta S=1$ 
terms in the Higgs Lagrangian density, calculated to zeroth order in  
the perturbation $\sigma^f$, which when extended to 
include the $\eta$ and pseudo Goldstone Higgs couplings reads
$$\eqalign{
{\cal L}_{\rm scnc}^{\Delta S=1}=&
\sum_{\xi=\phi,\eta}  \sum_{p=\pm} \sum_{F=R,I}
[\overline d \epsilon_{\xi F}^{(p)}(A_{\xi F12}^{(p)} + B_{\xi F12}^{(p)} 
\gamma_5)s+\overline s \epsilon_{\xi F}^{(p)} (A_{\xi F 21}^{(p)}
+B_{\xi F21}^{(p)}\gamma_5)d] \cr
+&
\overline d \epsilon_{\eta R}^{(3)}(A_{\eta R12}^{(3)} + B_{\eta R12}^{(3)} 
\gamma_5)s+\overline s \epsilon_{\eta R}^{(3)} (A_{\eta R21}^{(3)}
+B_{\eta R21}^{(3)}\gamma_5)d \cr
+&
\overline d \epsilon_{PG}^{(3)}(A_{PG 12} + B_{PG 12} 
\gamma_5)s+\overline s \epsilon_{PG }^{(3)} (A_{PG 21}
+B_{PG 21}\gamma_5)d~~~. \cr
}\eqno(8a)$$
The corresponding formula for the effective Hamiltonian density 
${\cal H}_{\rm eff}^{\Delta S=2}$ for the $\Delta S=2$ process $s+s \to d+d$ 
is   
$$\eqalign{{\cal H}_{\rm eff}^{\Delta S=2}
=&-{1\over 2}\sum_{\xi=\phi,\eta} \sum_{p=\pm} \sum_{F=R,I}
\overline d (A_{\xi F12}^{(p)} + B_{\xi F12}^{(p)} \gamma_5)s 
{1 \over M_{\xi F}^{2(p)} }
\overline d (A_{\xi F12}^{(p)} + B_{\xi F12}^{(p)} \gamma_5)s     \cr
-&{1\over 2}
\overline d (A_{\eta R12}^{(3)} + B_{\eta R12}^{(3)} \gamma_5)s 
{1 \over M_{\eta R}^{2(3)} }
\overline d (A_{\eta R12}^{(3)} + B_{\eta R12}^{(3)} \gamma_5)s     \cr
-&{1 \over 2}
\overline d (A_{PG12} + B_{PG12} \gamma_5)s {1 \over M_{PG}^2 }
\overline d (A_{PG12} + B_{PG12} \gamma_5)s~~~.     \cr
}\eqno(8b)$$
The eight Higgs squared masses appearing in Eq.~(8b) 
that carry superscripts $(\pm)$ are equal in pairs, 
$$M_{\xi R}^{2(+)}=M_{\xi I}^{2(-)}~,~~~
M_{\xi R}^{2(-)}=M_{\xi I}^{2(+)}~,~~~\xi=\phi,\eta~~~.\eqno(8c)$$
Although they (as well as $M_{\eta R}^{2(3)}$)
are given in terms of Lagrangian parameters by Eq.~(46b) 
and Table II of [1], we will not use these expressions, 
but rather will treat the  Higgs masses that are independent,  
after taking account of Eq.~(8c), directly as parameters to be 
bounded.  

The subscripts $12$ (or $21$) in Eq.~(8a) indicate 
the row 1 to column 2 (or row 2 to column 1) matrix 
element of the corresponding $2 \times 2$ matrix expressions for the 
$A$ and $B$ coefficients, which we now give.    
Because CP invariance of the Lagrangian for the six-Higgs doublet 
model implies that the Yukawa couplings appearing in Eq.~(8a) are
real, the $A$ coefficients appearing in Eq.~(8a) are related  
to the $B$ coefficients as follows, 
$$\eqalign{
A_{\phi,\eta R}^{(\pm)}=&-iB_{\phi,\eta I}^{(\pm)}~~~,\cr
A_{\phi,\eta I}^{(\pm)}=&iB_{\phi,\eta R}^{(\pm)}~~~, \cr      
A_{\eta R}^{(3)}=&i {\Omega_{AV} \over \Omega_{\phi} }B_{PG}~~~,\cr     
A_{PG}=&-i {\Omega_{\phi} \over \Omega_{AV}} B_{\eta R}^{(3)}~~~,\cr
\Omega_{AV}\equiv &(\Omega_{\phi}^2 + \Omega_{\eta}^2)^{1 \over 2}~~~.\cr
}\eqno(8d)$$
Defining 
$$
M_{2\times 2}^{(1)}=\pmatrix{1 & 0\cr 0 & 0 \cr}~,~~~    
M_{2\times 2}^{(2)}=\pmatrix{0 & 0\cr 0 & 1 \cr}~,~~~    
\rho_3=\pmatrix{1 & 0\cr 0 & -1 \cr}~~~,\eqno(9a)$$
the matrices $B_{\phi R,I}^{(\pm)} $ are given by
$$\eqalign{
B_{\phi R}^{(+)}=&{\surd 3 \over 4}g_{\phi}^d ({\hat V}_L^d {\hat 
V}_R^{d\dagger} 
-{\hat V}_R^d {\hat V}_L^{d\dagger})~~~,\cr
B_{\phi R}^{(-)}=&{\surd 3 \over 4}g_{\phi}^d ({\hat V}_L^d \rho_3 {\hat 
V}_R^{d\dagger} 
-{\hat V}_R^d \rho_3 {\hat V}_L^{d\dagger})~~~,\cr
B_{\phi I}^{(+)}=&{\surd 3 \over 4}g_{\phi}^d i ({\hat V}_L^d {\hat 
V}_R^{d\dagger} 
+{\hat V}_R^d {\hat V}_L^{d\dagger})~~~,\cr
B_{\phi I}^{(-)}=&{\surd 3 \over 4}g_{\phi}^d i ({\hat V}_L^d \rho_3 {\hat 
V}_R^{d\dagger} 
+{\hat V}_R^d \rho_3 {\hat V}_L^{d\dagger})~~~,\cr
}\eqno(9b)$$
the matrices $B_{\eta R,I}^{(\pm)}$ are given by 
$$\eqalign{
B_{\eta R}^{(+)}=-B_{\eta R}^{(-)}=&{\surd 3 \over 4}g_{\eta}^d
[\exp(i\theta) \hat {V}_L^d M_{2\times 2}^{(1)}{\hat V}_R^{d\dagger}
-\exp(-i\theta) {\hat V}_R^d M_{2\times 2}^{(1)}{\hat V}_L^{d\dagger}]~~~,\cr
B_{\eta I}^{(+)}=-B_{\eta I}^{(-)}=&{\surd 3 \over 4}g_{\eta}^d i
[\exp(i\theta) {\hat V}_L^d M_{2\times 2}^{(1)}{\hat V}_R^{d\dagger}
+\exp(-i\theta) {\hat V}_R^d M_{2\times 2}^{(1)}{\hat V}_L^{d\dagger}]~~~,\cr
}\eqno(9c)$$
and the matrices $B_{\eta R}^{(3)}$ and $B_{PG}$ are given by 
$$\eqalign{
B_{\eta R}^{(3)}=&{\surd 6 \over 4}  g_{\eta}^d 
[\exp(i\theta) {\hat V}_L^d M_{2 \times 2}^{(2)} {\hat V}_R^{d\dagger}-
 \exp(-i\theta){\hat V}_R^d M_{2 \times 2}^{(2)} {\hat V}_L^{d\dagger}]~~~,\cr
B_{PG}=&-{\surd 6 \over 4} i {\Omega_{\phi} \over \Omega_{AV} } 
g_{\eta}^d 
[\exp(i\theta){\hat V}_L^d M_{2 \times 2}^{(2)} {\hat V}_R^{d\dagger}+
 \exp(-i\theta){\hat V}_R^d M_{2 \times 2}^{(2)} {\hat V}_L^{d\dagger}]~~~.\cr
}\eqno(9d)$$  
In the above formulas, $\theta$ is the overall phase  
rotation angle between the $\phi$ and $\eta$ Higgs expectations 
introduced in Eq.~(21) of [1].

Taking the $K$ to $\overline K$ matrix element of Eq.~(8b), we get 
$$\langle K| {\cal H}_{\rm eff}^{\Delta S=2}|\overline K \rangle = 
-{1 \over 2} S_A^d \langle K| (\overline d s)^2 |\overline K \rangle 
-{1 \over 2} S_B^d \langle K| (\overline d \gamma_5 s)^2 
|\overline K \rangle~~~, 
\eqno(10a)$$
with 
$$\eqalign{
S_A^d=&\left(\sum_{\xi=\phi,\eta} \sum_{p=\pm}\sum_{F=R,I}
{(A_{\xi F\, 12}^{(p)})^2 \over M_{\xi F}^{2(p)} }\right) 
+{(A_{\eta R\, 12}^{(3)})^2 \over M_{\eta R}^{2(3)} }
+{A_{PG\,12}^2 \over M_{PG}^2 } ~~~,\cr
S_B^d=&\left(\sum_{\xi=\phi,\eta} \sum_{p=\pm}\sum_{F=R,I}
{(B_{\xi F\, 12}^{(p)})^2 \over M_{\xi F}^{2(p)} }\right) 
+{(B_{\eta R\, 12}^{(3)})^2 \over M_{\eta R}^{2(3)} }
+{B_{PG\,12}^2 \over M_{PG}^2 } ~~~.\cr
}\eqno(10b)$$
The corresponding formulas for the Higgs exchange contribution to the 
$D$ to $\overline D$ transition amplitude are obtained by replacing 
$K$ by $\overline D$, $d$ by $u$, and $s$ by $c$ in the above formulas, 
and replacing the explicit factors of $i$ by $-i$ in Eqs.~(9b, c, d),  
with the latter substitution reflecting the fact that the up sector  
Yukawa couplings involve the charge conjugates of the Higgs fields.  

Substituting now the explicit forms given in Eqs.~(7a-e) for the 
matrices ${\hat V}_{R,L}^{d,u}$, we get formulas for the 
sums $S_A^d$, $S_B^d$ that determine the Higgs exchange contribution to 
the $\overline K$ to $K$ transition amplitude, and for the 
corresponding sums $S_A^u$ and 
$S_B^u$ that contribute to the $D$ to $\overline D$ transition amplitude .  
With an eye 
to how these formulas will be used in Sec. IV, we write them as 
$$\eqalign{
S_A^d=&s_d^2c_d^2\exp(i\Phi_d)P_A^d~~~,\cr  
S_B^d=&-s_d^2c_d^2\exp(i\Phi_d)P_B^d~~~,\cr 
S_A^u=&s_u^2c_u^2\exp(i\Phi_u)P_A^u~~~,\cr  
S_B^u=&-s_u^2c_u^2\exp(i\Phi_u)P_B^u~~~,\cr  
}\eqno(11a)$$
with the positive real quantities $P_{A,B}^{d,u}$ given by
$$\eqalign{
P_A^d=&3\left[ (g_{\phi}^d)^2 \left( {\sin^2Y_d \over M_{\phi R}^{2(+)} }
+{\cos^2Y_d \over M_{\phi R}^{2(-)} } \right) \right.  \cr
+&\left.  {1\over 4} (g_{\eta}^d)^2 \left( {1 \over M_{\eta R}^{2(+)} }
+{1\over M_{\eta R}^{2(-)} } \right)
+{1 \over 2} (g_{\eta}^d)^2 \left( {\cos^2X_d \over M_{\eta}^{2(3)} }
+{\Omega_{\phi}^2 \over \Omega_{AV}^2 } 
{\sin^2X_d \over M_{PG}^2 } \right) \right]~~~,\cr
P_B^d=&3\left[ (g_{\phi}^d)^2 \left( {\cos^2Y_d \over M_{\phi R}^{2(+)} }
+{\sin^2Y_d \over M_{\phi R}^{2(-)} } \right) \right.  \cr
+&\left.  {1\over 4} (g_{\eta}^d)^2 \left( {1 \over M_{\eta R}^{2(+)} }
+{1\over M_{\eta R}^{2(-)} } \right)
+{1 \over 2} (g_{\eta}^d)^2 \left( {\sin^2X_d \over M_{\eta}^{2(3)} }
+{\Omega_{\phi}^2 \over \Omega_{AV}^2 } 
{\cos^2X_d \over M_{PG}^2 } \right) \right]~~~,\cr
P_A^u=&3\left[ (g_{\phi}^u)^2 \left( {\sin^2Y_u \over M_{\phi R}^{2(+)} }
+{\cos^2Y_u \over M_{\phi R}^{2(-)} } \right) \right.  \cr
+&\left.  {1\over 4} (g_{\eta}^u)^2 \left( {1 \over M_{\eta R}^{2(+)} }
+{1\over M_{\eta R}^{2(-)} } \right)
+{1 \over 2} (g_{\eta}^u)^2 \left( {\cos^2X_u \over M_{\eta}^{2(3)} }
+{\Omega_{\phi}^2 \over \Omega_{AV}^2 } 
{\sin^2X_u \over M_{PG}^2 } \right) \right]~~~,\cr
P_B^u=&3\left[ (g_{\phi}^u)^2 \left( {\cos^2Y_u \over M_{\phi R}^{2(+)} }
+{\sin^2Y_u\over M_{\phi R}^{2(-)} } \right) \right.  \cr
+&\left.  {1\over 4} (g_{\eta}^u)^2 \left( {1 \over M_{\eta R}^{2(+)} }
+{1\over M_{\eta R}^{2(-)} } \right)
+{1 \over 2} (g_{\eta}^u)^2 \left( {\sin^2X_u \over M_{\eta}^{2(3)} }
+{\Omega_{\phi}^2 \over \Omega_{AV}^2 } 
{\cos^2X_u \over M_{PG}^2 } \right) \right]~~~.\cr
}\eqno (11b)$$
The mixing and phase angles appearing in Eqs.~(11a, b) are given in terms  
of the various phase angles defined above by 
$$\eqalign{
Y_d=&{1\over 2}(\theta_1^d + \theta_2^d)~~~,\cr
X_d=&\theta+{1\over 2}(\phi_R^d-\phi_L^d)-{1\over 2}(\theta_1^d+\theta_2^d)~~~,
\cr
\Phi_d=&\theta_1^d-\theta_2^d-2\theta_{11}+2\theta_{21}-(\phi_R^d+\phi_L^d)~~~,
\cr
Y_u=&{1\over 2}(\theta_1^u+\theta_2^u)~~~,\cr
X_u=&-\theta+{1\over 2}(\phi_R^u-\phi_L^u)-{1\over 
2}(\theta_1^u+\theta_2^u)~~~,\cr
\Phi_u=&\theta_1^u-\theta_2^u-2\theta_{22}+2\theta_{21}-(\phi_R^u+\phi_L^u)~~~.
\cr
}\eqno(12a)$$

As a check on our phase conventions, we note 
that when the model is CP conserving, which implies [1] the 
additional condition $\sigma_{22}^f=\sigma_{11}^{f*}$, then Eq.~(4d) reduces   
to $\Theta=\pi/4$, and the following relations hold (modulo $\pi$), 
$$\eqalign{
\theta_1^{d,u}=&-\theta_2^{d,u}=\arg \sigma_{11}^{d,u}~~~,\cr
\phi_L^{d,u}=&-\arg \sigma_{12}^{d,u}-\arg \sigma_{11}^{d,u}~~~,\cr
\phi_R^{d,u}=&-\arg \sigma_{12}^{d,u}+\arg \sigma_{11}^{d,u}~~~,\cr
2\theta_{11}=&-2\theta_{22}=\phi_L^u-\phi_L^d~~~,\cr
2\theta_{12}=&-2\theta_{21}=-\phi_L^u-\phi_L^d~~~.\cr
}\eqno(12b)$$
When substituted into Eq.~(12a), these relations imply the vanishing 
(modulo $\pi$) of the 
phases $\Phi_d$ and $\Phi_u$. Consequently, with the phase 
conventions used in this paper, the 
imaginary parts of the Higgs exchange contributions to $K -\overline K$ and 
$\overline D-D$ mixing are a direct measure of the CP violating contributions 
to these   
amplitudes.

\bigskip
\centerline{\bf IV.~~Higgs Mass Bounds}

We proceed now to derive bounds on the Higgs masses in the six-Higgs   
doublet model.  Let 
$\Delta M_K^{\rm obs}$ and $\Delta M_D^{\rm obs}$ be respectively the 
measured value of the $K_1-K_2$ mass difference and the experimental 
upper bound on the $D_1-D_2$ mass difference.  Since it is reasonable 
to expect these to set upper limits on possible Higgs contributions to these  
mass differences, given respectively by [5]  
$$\eqalign{
|\Delta M_{K_1-K_2}^{\rm Higgs}| = & 
M_K^{-1} |\langle K|{\cal H}_{\rm eff}^{\Delta S=2} |\overline 
K\rangle|~~~,\cr
|\Delta M_{D_1-D_2}^{\rm Higgs}| = & 
M_D^{-1} |\langle \overline D|{\cal H}_{\rm eff}^{\Delta C=2} 
|D\rangle|~~~,\cr 
}\eqno(13a)$$
we get the basic inequalities 
$$\eqalign{
\Delta M_K^{\rm obs}\geq&M_K^{-1} |\langle K|{\cal H}_{\rm eff}^{\Delta S=2} 
|\overline K\rangle|~~~,\cr
\Delta M_D^{\rm obs}\geq&M_D^{-1} |\langle \overline D|
{\cal H}_{\rm eff}^{\Delta C=2} |D\rangle|~~~.\cr  
}\eqno(13b)$$
These inequalities will be used in this section, both independently and  
in combination, to derive a number of useful bounds on the Higgs masses. 

We begin by rewriting Eqs.~(10a, b) and (11a, b) so as to exhibit the 
features that play a role in our various inequalities.  Let us define   
$p_K$ and $p_D$ as the negatives of the ratios of the scalar to pseudoscalar 
matrix elements appearing in Eq.~(10a) and in its $D$ meson analog, 
$$\eqalign{
p_K=&-{\langle K| (\overline d s)^2 |\overline K \rangle   \over
\langle K| (\overline d \gamma_5 s)^2 |\overline K \rangle  }~~~,\cr
p_D=&-{\langle \overline D| (\overline u c)^2 |D \rangle   \over
\langle \overline D| (\overline u \gamma_5 c)^2 |D \rangle  }~~~.\cr
}\eqno(14a)$$
According to calculations of $p_K$ and $p_D$ by the vacuum 
insertion method [6, 7] 
and the MIT bag model [6], they are positive and small (roughly of order 
0.1 in magnitude).  The ratio $p_K$ can also be extracted from lattice   
calculations that have been performed [7] for kaon matrix elements, giving 
the result $p_K=0.30 \pm 0.05$, again of positive sign.  Although a similar 
lattice calculation is not yet available for the $D$ system, we will assume  
that this follows the same pattern as observed in the 
$K$ system, and that $p_D$  (as suggested by the vacuum saturation and 
bag model calculations) is positive.  
Substituting Eqs.~(10a), (11a), and (14a) into Eq.~(13b), our 
two basic inequalities now take the form 

$$\eqalign{
M_K\Delta M_K^{\rm obs}\geq&{1 \over 2} 
|\langle K|(\overline d \gamma_5 s)^2|\overline K\rangle|
s_d^2c_d^2 |P_B^d+p_KP_A^d|~~~,\cr
M_D\Delta M_D^{\rm obs}\geq&{1 \over 2} 
|\langle \overline D|(\overline u \gamma_5 c)^2|D\rangle|
s_u^2c_u^2 |P_B^u+p_DP_A^u|~~~, \cr  
}\eqno(14b)$$
with $P_B^{d,u}+p_{K,D}P_A^{d,u}$  both sums of positive terms.  
Introducing the definitions 
$$\eqalign{
E_K=&{2 M_K\Delta M_K^{\rm obs} \over s_{12}^2 c_{12}^2 
|\langle K|(\overline d \gamma_5 s)^2|\overline K\rangle| }~~~,\cr
E_D=&{2 M_D\Delta M_D^{\rm obs} \over s_{12}^2 c_{12}^2 
|\langle \overline D|(\overline u \gamma_5 c)^2|D\rangle| }~~~,\cr
}\eqno(14c)$$
we rewrite the inequalities of Eq.~(14b) as
$$\eqalign{
\left({E_K \over |P_B^d+p_K P_A^d|} \right)^{1\over 2} \geq &
{|s_d c_d| \over s_{12} c_{12} } ~~~,\cr
\left({E_D \over |P_B^u+p_D P_A^u|} \right)^{1\over 2} \geq &
{|s_u c_u| \over s_{12} c_{12} } ~~~.\cr
}\eqno(14d)$$

Although Eqs. (14d) are relevant for the numerical fits of Sec. VI, where 
the products $|s_d c_d|$ and $|s_u c_u|$ are known, they cannot be used 
to give fit-independent bounds on the Higgs masses, 
because either  $|s_d c_d|$ or $|s_u c_u|$  
can vanish.  However, we shall now show that the sum 
$|s_dc_d|+|s_uc_u|$ is bounded below by CKM matrix elements, 
permitting us to extract a useful inequality by combining  
the $K$ meson and $D$ meson flavor changing neutral current constraints.  
To see this, we substitute Eq.~(4b) for $V_L^{d,u}$, together  
with the definitions of Eq.~(7e), into Eq.~(6b) for the adjoint of the 
unrephased CKM
matrix, and take 
absolute values of the matrix elements on the first row, giving 
$$\eqalign{
s_{12}=&|c_us_d\exp(-i\phi_L^d)-s_uc_d\exp(-i\phi_L^u)|
\leq |c_us_d|+|s_uc_d|~~~,\cr
c_{12}=&|c_uc_d+\exp(i\phi_L^u-i\phi_L^d)s_us_d|
\leq |c_uc_d|+|s_us_d|~~~.\cr
}\eqno(15a)$$
Multiplying these inequalities, we get 
$$s_{12}c_{12}\leq|c_ds_d|(s_u^2+c_u^2)+|c_us_u|(s_d^2+c_d^2)
=|c_ds_d|+|c_us_u|~~~,\eqno(15b)$$
giving the needed lower bound.  
Adding the two inequalities in Eq.~(14d), and using Eq.~(15b),  
 we get the master inequality
$$\left({E_K \over |P_B^d+p_KP_A^d|} \right)^{1\over 2}
+
\left({E_D \over |P_B^u+p_DP_A^u|} \right)^{1\over 2}
\geq 1 ~~~.\eqno(16)$$

Since all terms in the denominators 
$|P_B^{d,u}+p_{K,D}P_A^{d,u}|$ are positive, deleting 
any of these terms serves to make the left hand side of Eq.~(16) larger, 
giving a number of simpler subsidiary inequalities that are consequences of 
the master inequality.   Specifically, if we delete all terms in both  
denominators that do not refer to a given Higgs mass (i.e., if we set all 
of the other Higgs masses equal to infinity), we get a lower bound for 
the Higgs mass that we have retained; performing this in succession for 
the six Higgs masses we get the following inequalities,
$$\eqalign{
M_{\phi R}^{(+)} \geq& \left\{
\left[ {E_K\over 3 (g_{\phi}^d)^2 (\cos^2Y_d+p_K \sin^2 Y_d) } 
\right]^{1\over 2}   
+\left[ {E_D\over 3 (g_{\phi}^u)^2 (\cos^2Y_u+p_D \sin^2 Y_u) } 
\right]^{1\over 2} \right\}^{-1}~~~,\cr  
M_{\phi R}^{(-)} \geq& \left\{
\left[ {E_K\over 3 (g_{\phi}^d)^2 (\sin^2Y_d+p_K \cos^2 Y_d) } 
\right]^{1\over 2}   
+\left[ {E_D\over 3 (g_{\phi}^u)^2 (\sin^2Y_u+p_D \cos^2 Y_u) } 
\right]^{1\over 2} \right\}^{-1}~~~,\cr  
M_{\eta R}^{(\pm)} \geq& \left\{
\left[ {4E_K\over 3 (g_{\eta}^d)^2 (1+p_K) } 
\right]^{1\over 2}   
+\left[ {4E_D\over 3 (g_{\eta}^u)^2 (1+p_D) } 
\right]^{1\over 2} \right\}^{-1}~~~,\cr  
M_{\eta R}^{(3)} \geq& \left\{
\left[ {2E_K\over 3 (g_{\eta}^d)^2 (\sin^2X_d+p_K \cos^2 X_d) } 
\right]^{1\over 2}   
+\left[ {2E_D\over 3 (g_{\eta}^u)^2 (\sin^2X_u+p_D \cos^2 X_u) } 
\right]^{1\over 2} \right\}^{-1}~~~,\cr  
M_{PG} \geq& \left\{
\left[ {2E_K\Omega_{AV}^2\over 3 (g_{\eta}^d)^2 \Omega_{\phi}^2
(\cos^2X_d+p_K \sin^2 X_d) } 
\right]^{1\over 2}   
+\left[ {2E_D\Omega_{AV}^2\over 3 (g_{\eta}^u)^2 \Omega_{\phi}^2
(\cos^2X_u+p_D \sin^2 X_u) } 
\right]^{1\over 2} \right\}^{-1}~.\cr  
}\eqno(17a)$$
Applying the same procedure of successive deletion of denominator terms 
to the inequalities of Eq.~(14d), we get a set of analogous inequalities 
[which, by use of Eq.~(15b), imply those of Eq.~(17a)] that will be used 
in the numerical work of Sec. VI, 
$$\eqalign{
M_{\phi R}^{(+)} \geq&\max  \left\{ r_d
\left[ {E_K\over 3 (g_{\phi}^d)^2 (\cos^2Y_d+p_K \sin^2 Y_d) } 
\right]^{-{1\over 2}}   
,r_u\left[ {E_D\over 3 (g_{\phi}^u)^2 (\cos^2Y_u+p_D \sin^2 Y_u) } 
\right]^{-{1\over 2}} \right\}~~~,\cr  
M_{\phi R}^{(-)} \geq&\max  \left\{ r_d
\left[ {E_K\over 3 (g_{\phi}^d)^2 (\sin^2Y_d+p_K \cos^2 Y_d) } 
\right]^{-{1\over 2}}   
,r_u\left[ {E_D\over 3 (g_{\phi}^u)^2 (\sin^2Y_u+p_D \cos^2 Y_u) } 
\right]^{-{1\over 2}} \right\}~~~,\cr  
M_{\eta R}^{(\pm)} \geq&\max  \left\{ r_d
\left[ {4E_K\over 3 (g_{\eta}^d)^2 (1+p_K) } 
\right]^{-{1\over 2}}   
,r_u \left[ {4E_D\over 3 (g_{\eta}^u)^2 (1+p_D) } 
\right]^{-{1\over 2}} \right\}~~~,\cr  
M_{\eta R}^{(3)} \geq&\max  \left\{ r_d
\left[ {2E_K\over 3 (g_{\eta}^d)^2 (\sin^2X_d+p_K \cos^2 X_d) } 
\right]^{-{1\over 2}}   
,r_u \left[ {2E_D\over 3 (g_{\eta}^u)^2 (\sin^2X_u+p_D \cos^2 X_u) } 
\right]^{-{1\over 2}} \right\}~~~,\cr  
M_{PG} \geq&\max  \left\{ r_d
\left[ {2E_K\Omega_{AV}^2\over 3 (g_{\eta}^d)^2 \Omega_{\phi}^2
(\cos^2X_d+p_K \sin^2 X_d) } 
\right]^{-{1\over 2}}   
,r_u \left[ {2E_D\Omega_{AV}^2\over 3 (g_{\eta}^u)^2 \Omega_{\phi}^2
(\cos^2X_u+p_D \sin^2 X_u) } 
\right]^{-{1\over 2}} \right\}~~~,\cr  
r_d\equiv& {|s_d c_d| \over s_{12}c_{12}}    
~,~~~r_u\equiv {|s_u c_u|\over s_{12}c_{12}} ~.\cr
}\eqno(17b)$$

The bounds in Eq.~(17a) still depend on the mixing angles $X_{d,u}$ and 
$Y_{d,u}$ defined in Eq.~(12a); a set of (necessarily weaker) bounds that 
do not depend on these angles is obtained by using the inequalities, 
valid for $p\leq 1$, 
$$\eqalign{
\cos^2 Z + p \sin^2 Z=&(1-p) \cos^2 Z + p \geq p~~~,\cr     
\sin^2 Z + p \cos^2 Z=&(1-p) \sin^2 Z + p \geq p~~~,\cr     
}\eqno(18a)$$
giving (for $p_{K,D}\leq 1$) the inequalities 
$$\eqalign{
M_{\phi R}^{(\pm)}&\geq \left\{ 
\left[ {E_K \over 3 (g_{\phi}^d)^2 p_K } \right]^{1 \over 2}     
+\left[ {E_D \over 3 (g_{\phi}^u)^2 p_D } \right]^{1 \over 2}     
\right\}^{-1}~~~,\cr
M_{\eta R}^{(\pm)}&\geq \left\{ 
\left[ {4E_K \over 3 (g_{\eta}^d)^2 (1+p_K) } \right]^{1 \over 2}     
+\left[ {4E_D \over 3 (g_{\eta}^u)^2 (1+p_D) } \right]^{1 \over 2}     
\right\}^{-1}~~~,\cr
M_{\eta R}^{(3)}&\geq \left\{ 
\left[ {2E_K \over 3 (g_{\eta}^d)^2 p_K } \right]^{1 \over 2}     
+\left[ {2E_D \over 3 (g_{\eta}^u)^2 p_D } \right]^{1 \over 2}     
\right\}^{-1}~~~,\cr
M_{PG}&\geq \left\{ 
\left[ {2E_K \Omega_{AV}^2 \over 
3 (g_{\eta}^d)^2 \Omega_{\phi}^2 p_K } \right]^{1 \over 2}     
+\left[ {2E_D \Omega_{AV}^2 \over 
3 (g_{\eta}^u)^2 \Omega_{\phi}^2 p_D } \right]^{1 \over 2}     
\right\}^{-1}~~~.\cr
}\eqno(18b)$$
Equations (17a, b) and (18b) are our final bounds for the Higgs 
masses in the six-Higgs doublet model.  

To obtain numerical values from the bounds of Eq.~(18b), we first  
need to evaluate the ratios $E_{K,D}$ defined in Eq.~(14c).  For $E_K$ 
we use the measured value [8] 
$\Delta M_K^{\rm obs}=3.49 \times 10^{-12}~ {\rm MeV}$, 
together with $M_K=497.7~ {\rm MeV}$,~$M_d=6~ {\rm MeV}$,
~$M_s=115~ {\rm MeV}$,~$f_K=160~ {\rm MeV}$, $c_{12}=0.975$,  
$s_{12}=0.221$, and the lattice evaluation [7] 
$$|\langle K |(\overline d \gamma_5 s)^2 | \overline K \rangle|
=1.58 \left( {M_K \over M_s +M_d} \right)^2 M_K^2 f_K^2~~~,\eqno(19a)$$ 
to give 
$$E_K=0.44\times 10^{-12}  ({\rm GeV})^{-2}~~~.\eqno(19b)$$   
For $E_D$, we use the experimental upper bound [8] 
$\Delta M_D^{\rm obs}=1.58 \times 10^{-10}~ {\rm MeV}$, together with 
$M_D=1865~ {\rm MeV}$,~$M_u=3.25~ {\rm MeV}$,~$M_c=1.25~ {\rm GeV}$,  the  
lattice calculation [9] value $f_D\simeq 1.2 f_K$, 
and the vacuum saturation approximation 
formula [7] 
$${ |\langle K| (\overline d \gamma_5 s)^2 | \overline K \rangle \over 
|\langle \overline D| (\overline u \gamma_5 c)^2 | D \rangle| }
={f_K^2 M_K^2 [11  M_K^2 (M_s+M_d)^{-2} +1] \over 
 f_D^2 M_D^2 [11 M_D^2 (M_c+M_u)^{-2} +1] }~~~,
 \eqno(19c)$$ 
 to give 
$$E_D=27\times 10^{-12}  ({\rm GeV})^{-2}~~~.\eqno(19d)$$   
Since the scalar to pseudoscalar ratio $p_D$ has not yet been computed on 
the lattice, we will assume that $p_D=p_K=0.3$ in evaluating Eqs.~(18b). 

To complete the computation of Higgs mass bounds, we need the values of 
the various Yukawa couplings appearing in Eqs.~(18b).  Here some 
assumptions about how the first and second family masses are generated 
are needed.  If, following [1], we assume that the $\phi$ Higgs expectations 
generate the third family masses, the $\eta$ Higgs expectations generate 
the second family masses, while cyclic asymmetries in the Yukawa couplings  
are responsible for the first family masses, then we get from Eq.~(32b) 
of [1] the formulas 
$$\eqalign{
g_{\phi}^u=&{M_t \over 3 \Omega_{\phi} }~,~~ 
g_{\phi}^d={M_b \over 3 \Omega_{\phi} }~~~,\cr
g_{\eta}^u=&{M_c \over 3 \Omega_{\eta} }~,~~ 
g_{\eta}^d={M_s \over 3 \Omega_{\eta} }~~~.\cr
}\eqno(20a)$$
If we now assume equal $\Omega_{\phi}$ and $\Omega_{\eta}$, so that  
$\Omega_{\phi}=\Omega_{\eta}=71~ {\rm GeV}$, we get the numerical values 
$$\eqalign{
g_{\phi}^u=&0.82~,~~g_{\phi}^d=0.020~~~,\cr 
g_{\eta}^u=&0.0059~,~~g_{\eta}^d=0.00054~~~,\cr
}\eqno(20b)$$
giving the Higgs mass lower bounds 
$$\eqalign{
M_{\phi R}^{(\pm)} \geq& 24~ {\rm TeV}~~~,\cr 
M_{\eta R}^{(\pm)} \geq& 470~ {\rm GeV}~~~,\cr
M_{\eta R}^{(3)} \geq& 320~ {\rm GeV}~~~,\cr
M_{PG} \geq& 220~ {\rm GeV}~~~.
}\eqno(20c)$$

An alternative possibility, discussed in the next section, is that the 
second family masses are generated by cyclic asymmetries in the $\phi$ 
Yukawa couplings, with the first family masses generated by the $\eta$ 
Higgs expectations.  In this case, the second line of Eq.~(20a) is 
replaced by 
$$g_{\eta}^u\simeq {M_u \over 1.5 \Omega_{\eta} }~,~~ 
g_{\eta}^d\simeq {M_d \over 1.5 \Omega_{\eta} }~~~,\eqno(21a)$$
giving (for $\Omega_{\eta}=\Omega_{\phi}$) the numerical 
values 
$g_{\eta}^u=3.1 \times 10^{-5}~,~~g_{\eta}^d=5.7 \times 10^{-5}~~~,$
which imply the much weaker $\eta$ and pseudo Goldstone Higgs mass bounds 
$$\eqalign{
M_{\eta R}^{(\pm)} \geq& 5.5~ {\rm GeV}~~~,\cr
M_{\eta R}^{(3)} \geq& 3.7~ {\rm GeV}~~~,\cr
M_{PG} \geq& 2.6~ {\rm GeV}~~~.
}\eqno(21b)$$

Finally, we note that the first inequality of Eq.~(18b) also applies to 
the CP conserving case of the three-Higgs doublet model of [1], for which  
$\Omega_{\phi}=\surd 2 \times 71~{\rm GeV}$, so that $g_{\phi}^{d,u}$ are a 
factor of $\surd 2$ smaller than given in Eq.~(20b).  This reduces the 
corresponding bounds of Eq.~(18b) by a factor of $\surd 2$, giving for 
the CP-conserving three-Higgs doublet model the Higgs mass lower bounds 
$$ M_{\phi R}^{(\pm)} \geq 17~ {\rm TeV}~~~.\eqno(22)$$
In this model there are no $\eta$ Higgs states, and hence no possibility of   
neutral Higgs states that are not supermassive.  
\bigskip
\centerline{\bf V.~~Implications of Requiring Naturally Small First 
Family Masses}

We saw in the preceding section that the bounds on the $\eta$  and 
pseudo Goldstone Higgs masses depend on the coupling pattern assumed 
for the $\eta$ Higgs discrete chiral triplet.  In this section we 
classify possible $\eta$ Higgs coupling patterns, based on a criterion  
of requiring naturally small first family masses.  Referring to Eqs.~(1a) 
and (1b), we see that contributions to the mass matrix in the six-Higgs 
doublet 
model are of three distinct types, arising from deviations from cyclic 
symmetry in the $\phi$ Higgs Yukawa couplings, deviations from cyclic 
symmetry in the $\phi$ Higgs expectations, and contributions from the 
weakly coupled $\eta$ Higgs expectations.  Since these three 
contributions are not directly related physically, detailed cancellations 
between them in the determination of the first family masses are 
{\it a priori} unlikely. Hence as a necessary (but not sufficient) 
condition for naturally small first family masses, we impose the condition 
that only one of these three contributions dominates in a 
leading approximation 
in which the first family masses are exactly zero.

We begin by noting that the deviations $\delta_n^f$ from cyclic symmetry in 
the $\phi$ Higgs Yukawa couplings cannot dominate and 
lead to naturally zero first family masses.  Let us suppose that the 
$\delta_n^f$ do dominate, and consider first the case in 
which the model chooses a CP conserving ground state, for which the 
parameters $\delta_n^f$ are all real.  In this case the magnitudes 
$$\eqalign{
|\sigma_{11}^f|\simeq&|\delta_3^f+\overline \omega \delta_2^f +\omega 
\delta_1^f|~~~,\cr    
|\sigma_{22}^f|\simeq&|\delta_3^f+\omega \delta_2^f +\overline \omega 
\delta_1^f|~~~,\cr    
}\eqno(23)$$
are equal, and so the first and second family masses are equal.  Turning 
on a CP violation results in complex $\delta_n^f$'s, for which the 
first and second family masses are no longer the same, but clearly a 
fine tuning of the amount of CP violation would be needed to achieve 
zero first family masses.   Hence dominance of the Higgs expectation 
asymmetries $\delta_n^f$ is not compatible with naturally small first 
family masses.

We consider next the case in which the contribution $R^f$ of the $\eta$ 
Higgs expectations dominates, which is the scenario assumed in [1].  In this 
case the $\eta$ Higgs expectations give rise to the second family masses, 
and the leading approximation to the first family masses is automatically 
zero, satisfying our criterion for naturally small first family masses.  
However, a potential problem arises when we examine the structure of 
the CKM matrix.  In the leading approximation in which only $R^f$ is 
retained in the mass matrix, the CKM matrix is unity.  To get a nontrivial 
CKM matrix, we must add back the small perturbations $\delta_n^f$ and 
$\mu_{\ell m}^f$ in Eq.~(1b).  According to Eqs.~(4a-d), in each flavor 
channel we then get 
$$\eqalign{
A-B \simeq & -|\sigma_{22}|^2~~~,\cr
|z| \simeq & |\sigma_{12}||\sigma_{22}|~~~,\cr
\Theta \simeq & {1 \over 2} \tan^{-1} \left( {2 |\sigma_{12}| 
\over |\sigma_{22}|} \right) ~~~,\cr
}\eqno(24a)$$
which by the hypothesis of dominance of $R^f$ is much less than unity. 
Hence, in particular, the up channel quantity  $s^u=\sin \Theta^u$ is 
much less than unity in magnitude.  But referring now to the corrected 
expressions [3] 
for the CKM elements $s_{13}$ and $s_{23}$, we have 
$$\eqalign{
s_{13}=&|s_3-d_3|/3~,~~~s_{23}=|s_3+d_3|/3~~~,\cr
s_3=&c_u(\sigma_{13}^d-\sigma_{13}^u)~,
~~~d_3=s_u\exp(-i\phi_L^u) (\sigma_{23}^d-\sigma_{23}^u)~~~,\cr
}\eqno(24b)$$
indicating that the spread of $s_{23}$ and $s_{13}$ from their geometric 
mean is suppressed by the small quantity $s_u$.  This in turn requires 
relatively large parameter values $\sigma_{23}$ and/or $\sigma_{13}$ 
to give a satisfactory fit to the data, contradicting the starting assumption 
of a dominant $R^f$.  We shall see evidence for this phenomenon in the 
next section, where we find Yukawa asymmetries comparable in 
magnitude to $R^f$, and hence substantial fine tuning in 
achieving small first family masses.  

We turn finally to the third case, in which the 
dominant contributions to the 
mass matrix come from the asymmetries $\mu_{\ell m}^f$ 
of the $\phi$ Higgs Yukawa couplings.   For the leading approximation 
to the $2 \times 2$ submatrix $m$ of the mass matrix, we then have 
(suppressing the flavor index $f$) 
$$m={1\over 3} \pmatrix{ \mu_{11} & \mu_{12} \cr 
\mu_{21} & \mu_{22} \cr}~~~,\eqno(25a)$$
with only two of the matrix elements in Eq.~(25a) independent, since  
CP invariance of the $\phi$ Higgs Yukawa couplings implies [1] that 
$$\mu_{21}=\mu_{12}^*~,~~~\mu_{22}=\mu_{11}^*~~~.\eqno(25b)$$
In order for Eq.~(25a) to have a zero eigenvalue, we must impose the 
additional condition 
$$|\mu_{12}|=|\mu_{11}| \Leftrightarrow  \mu_{12}=\exp(-i\chi) \mu_{11}^*
~~~,\eqno(26a)$$ 
an explanation for which must be sought in  
higher energy physics determining the Yukawa couplings.  
Taken together, Eqs.~(26a) and (25b) imply that 
the matrix $m$ takes the rank one form 
$$m={1\over 3} \pmatrix{ \mu_{11} & \exp(-i\chi)\mu_{11}^* \cr 
\exp(i\chi) \mu_{11} & \mu_{11}^* \cr}~~~,\eqno(26b)$$
with eigenvalues $|\kappa_1|=0$ and $|\kappa_2|={2 \over 3} |\mu_{11}|$, 
corresponding  respectively to the first and second family mass eigenstates.  
>From 
Eq.~(4a), we find that the diagonalizing matrices $V_{L,R}$ are given 
by Eq.~(4b), with 
$$\eqalign{
\Theta =&{\pi \over 4}~,~~~\cos\Theta= \sin\Theta={1 \over \surd 2}~~~,\cr
\phi_L=& \chi~,~~~\phi_R=\chi+2~ \arg \mu_{11}~~~.\cr
}\eqno(27a)$$
Referring to Eq.~(15a), we see that the sine of the Cabibbo angle 
$s_{12}$ is given now by  
$$s_{12}={1\over 2}|\exp(-i\phi_L^d)-\exp(-i\phi_L^u)|
=|\sin{1\over 2}(\chi^u-\chi^d)|~~~.\eqno(27b)$$
Averaging Eq.~(24b) and the analogous expression obtained  
from the lower left corner of the CKM matrix, we get the following 
leading order expressions for $s_{13}$ and $s_{23}$,  
$$\eqalign{
s_{13}=&|s_3-d_3|/3~,~~~s_{23}=|s_3+d_3|/3~~~,\cr
s_3=&{1 \over \surd 2}(\sigma_{13}^d-\sigma_{13}^u)~,
~~~d_3={1 \over \surd 2}\exp[-{i\over 2}(\chi^u+\chi^d)] \cos[{1 \over 2}
(\chi^u-\chi^d)](\sigma_{23}^d-\sigma_{23}^u)~~~,\cr
}\eqno(27c)$$
in which the coefficient of $\sigma_{23}^d-\sigma_{23}^u$ in $d_3$  
is not now a small parameter.  

To complete the analysis of the third case, let us calculate the 
first family mass eigenvalue.  There are four possible contributions 
to a nonzero first family mass: (i) deviations from the rank one condition 
of Eq.~(26a) on the $2 \times 2$ submatrix $m$ of the mass matrix, (ii) 
 asymmetries in the 
$\phi$ Higgs expectations $\delta_n$, (iii) couplings to the third family 
through the mass matrix elements  
$\sigma_{13},~\sigma_{31}$  and $\sigma_{23},~\sigma_{32}$, and 
(iv) effects of the 
$\eta$ Higgs expectation term $R$ in Eq.~(1b).  (We are continuing to 
suppress the flavor index $f$ when not needed.)  
The simplest way to calculate the   
first family mass matrix eigenvalue $|\kappa_1|$ is to 
evaluate the absolute value of the determinant of the $3 \times 3$ mass 
matrix $M^{\prime}$ of Eq.~(2a), which yields $|\kappa_1|$ when divided 
by the product of the other two eigenvalues, giving to leading order in 
small quantities, 
$$|\kappa_1|\simeq {|\det M^{\prime}|\over 3 |\kappa_2| }
\simeq {|\det(3M^{(3)}+\sigma)| \over 2 |\mu_{11}|}~~~.\eqno(28a)$$
To illustrate this in a simple texture model corresponding to the third 
case, let us assume that the contributions (i) and (ii) above are absent, 
that is, we assume that the condition of Eq.~(26a) is exact and that 
cyclic asymmetries in the $\phi$ Higgs expectations are negligible.  Then 
the only relevant contributions to the determinant in Eq.~(28a) are (iii) 
and (iv) above, and a simple calculation gives the leading order formula
$$|\kappa_1|\simeq
|{3 \over 2} R-{1 \over 6}[\sigma_{13}^* - \exp(i\phi_L)\sigma_{13}]
[\sigma_{31}^* - \exp(-i\phi_R)\sigma_{31}]|~~~,\eqno(28b)$$ 
with $\phi_{L,R}$ given in Eq.~(27a). When only the $R$ term is  
retained, substituting Eq.~(28b) into Eq.~(1a) yields 
the formulas of Eq.~(21a).  Within the simplified texture model, we have also 
calculated 
the CP violating angle $\delta_{13}$ appearing in the standard form [8] 
CKM matrix as a consequence of the CP violation carried by the $R$ term.  
After considerable algebra, we find
$$\delta_{13}\simeq 2\surd 2 \epsilon^u[\cot(\phi_L^u-\phi_L^d) 
-\cot(\phi_L^u+2\theta_{13})]-2\surd 2 \epsilon^d \csc(\phi_L^u-\phi_L^d)~~~,
\eqno(29a)$$
with the auxiliary quantities appearing in Eq.~(29a) defined by 
$$\theta_{13}={\rm arg}(\sigma_{13}^d-\sigma_{13}^u)~,~~~
\epsilon^{u,d}={-9 {\rm Re}(\mu_{11}^{u,d}R^{u,d}) \over
4 \surd 2 |\mu_{11}^{u,d}|^2 }~~~.\eqno(29b)$$

To complete the specification of the texture model corresponding to the 
third case, we note that since the $2 \times 2$ 
diagonalizing submatrices $V_{L,R}$
are maximally mixing in this case, wherever ``1st'' or ``2nd'' appears in 
the Higgs meson fermion family coupling Table II of [1], there now should 
appear ``1st and 2nd'', indicating couplings of equal magnitude 
of first family to 
first family, first family to second family, and second family to second 
family.   We also remark that the rank one condition of Eq.~(26a) 
can be reexpressed as a model for the Yukawa asymmetries  
$\beta_{\ell m}$, by using the inversion formulas 
$$\eqalign{
\beta_{11}=&{2 \over 9}{\rm Re}(\mu_{11}+\mu_{12}+\mu_{13}+\mu_{31})~~~,\cr 
\beta_{12}=&{2 \over 9}{\rm Re}(\omega\mu_{11}+\overline \omega \mu_{12}
+\mu_{13}+\omega\mu_{31})~~~,\cr 
\beta_{13}=&{2 \over 9}{\rm Re}(\overline \omega \mu_{11}+\omega \mu_{12}
+\mu_{13}+\overline \omega \mu_{31})~~~,\cr 
\beta_{21}=&{2 \over 9}{\rm Re}(\omega \mu_{11}+\omega \mu_{12}
+\omega \mu_{13}+\mu_{31})~~~,\cr 
\beta_{22}=&{2 \over 9}{\rm Re}(\overline \omega \mu_{11}+\mu_{12}
+\omega \mu_{13}+\omega \mu_{31})~~~,\cr 
\beta_{23}=&{2 \over 9}{\rm Re}(\mu_{11}+\overline \omega \mu_{12}
+\omega \mu_{13}+\overline \omega \mu_{31})~~~,\cr 
\beta_{31}=&{2 \over 9}{\rm Re}(\overline \omega \mu_{11}+\overline \omega 
\mu_{12}+\overline \omega \mu_{13}+\mu_{31})~~~,\cr 
\beta_{32}=&{2 \over 9}{\rm Re}(\mu_{11}+\omega \mu_{12}
+\overline \omega \mu_{13}+\omega \mu_{31})~~~,\cr 
\beta_{33}=&{2 \over 9}{\rm Re}(\omega \mu_{11}+\mu_{12}
+\overline \omega \mu_{13}+\overline \omega \mu_{31})~~~.\cr 
}\eqno(30)$$

\bigskip
\centerline{\bf VI.~~Some Illustrative Numerical Fits}

We give in this section illustrative numerical fits, obtained by 
the method of minimizing a ``cost function'' described in 
detail in Sec. IX of [1]. The mass and CKM cost functions are as in Eqs.  
(51b) and (52b) of [1], with the changes that we omit the term 
$(\Omega_{\phi}-\Omega_{\eta})^2$, which had little effect on the fits, 
and take the target values and standard deviations 
from the latest Particle Data Group [8] 
compilation.  For the parameter cost function, we use Eq.~(53a) of [1],   
with the changes that we omit the summation restrictions excluding 
the $n=3$ and $m=n=3$ terms, 
so that all asymmetries are treated symmetrically,  and we
take the exponent $\epsilon$ as 2 rather than as 1.  For the third case 
of the proceeding section, in which the second family masses arise from 
Yukawa coupling asymmetries, we also add to the parameter cost function 
a term 
$${\Omega_{\eta}^2 \over 4} \left[ \left( {g_{\eta}^u \over M_u} \right)^2 
+\left( {g_{\eta}^d \over M_d} \right)^2   
+\left( {g_{\eta}^e \over M_e} \right)^2  \right]~~~,\eqno(31)$$
designed to keep the $\eta$ Higgs contributions small, and start the 
iteration from preliminary parameter values determined with $g_{\eta}^f
=0,~f=u,d,e$.   
We omit flavor changing neutral current constraints from the cost function, 
so that there is no fine tuning to attempt to suppress flavor 
changing neutral current effects;  instead we use the inequalities 
of Eq.~(17b), evaluated using the parameters determined by the 
fitting procedure, to give lower bounds on the Higgs masses that guarantee 
sufficiently small flavor changing neutral current contributions to 
$K-\overline K$ and $\overline D -D$ mixing.  

Fitting results for the second case of the proceeding section are given 
in Table I, obtained with a standard deviation for the Yukawa asymmetries 
$\beta$ of $\sigma_{\rm parameter}=0.02$.  For this fit, the maximum $|\beta|$ 
values in the up, down, and electron sectors are 0.032, 0.058, 
and 0.010 respectively.  Although the iteration is started with the $\eta$ 
Higgs coupling values of Eq.~(20b), the converged fit has significantly 
smaller couplings $g_{\eta}^u=0.0031$ 
and $g_{\eta}^d=0.00039$, indicating that the Yukawa asymmetries make 
substantial contributions to the second family masses.  This means that  
the $\eta$ Higgs contributions do not dominate the Yukawa asymmetries, 
and thus substantial fine tuning is involved in achieving small 
first family masses.  For the fit of Table I, the Higgs mass bounds obtained 
from Eq.~(17b) are
$$\eqalign{
M_{\phi R}^{(+)}\geq&520~{\rm TeV}              ~~~,\cr
M_{\phi R}^{(-)}\geq&440~{\rm TeV}              ~~~,\cr 
M_{\eta R}^{(\pm)}\geq&1.3~{\rm TeV}              ~~~,\cr 
M_{\eta R}^{(3)}\geq&1.1~{\rm TeV}              ~~~,\cr 
M_{PG}\geq&1.0~{\rm TeV}              ~~~.\cr 
}\eqno(32a)$$

In Table II we give fitting results for the third case of the preceding 
section, obtained now with a standard deviation for the Yukawa asymmetries 
$\beta$ of $\sigma_{\rm parameter}=0.08$.  For this fit, the maximum $|\beta|$ 
values in the up, down, and electron sectors are 0.042, 0.052, 
and 0.11 respectively, with the relatively large Yukawa coupling asymmetry 
needed in the electron 
sector reflecting the fact that in this sector the ratio of the second  
family to third family mass is biggest.  For the fit of Table II, the   
Higgs mass bounds obtained from Eq.~(17b) are
$$\eqalign{
M_{\phi R}^{(+)}\geq&370~{\rm TeV}              ~~~,\cr
M_{\phi R}^{(-)}\geq&620~{\rm TeV}              ~~~,\cr 
M_{\eta R}^{(\pm)}\geq&210~{\rm GeV}              ~~~,\cr 
M_{\eta R}^{(3)}\geq&220~{\rm GeV}              ~~~,\cr 
M_{PG}\geq&140~{\rm GeV}              ~~~.\cr 
}\eqno(32b)$$

To conclude, in order for flavor changing neutral current effects in our 
models to be sufficiently small, the $\phi$ Higgs masses must be very large,
far outside the regime in which conventional perturbative Higgs physics 
applies (see [10] for a recent review of both perturbative 
and strongly coupled Higgs models).  Our results are consistent with 
general group theoretic analyses of flavor changing neutral currents 
in multi-Higgs doublet extensions of the standard model [11], which when 
applied to our models imply that flavor changing neutral currents cannot 
cancel kinematically, but must be eliminated either by fine tuning (an 
option we have ruled out by the inequalities of Sec.~IV) or by having some 
very large Higgs masses.  From an experimental viewpoint, the most 
interesting scenario within our models corresponds to the third case 
discussed in Sec.~V, in which the $\eta$ Higgs couplings are small enough 
that their expectations play an important role only in determining the 
first family masses and in giving rise to CP violation.      
In this case the $\eta$ Higgs and pseudo Goldstone 
Higgs states are permitted by our mass bounds to be 
light enough to be seen in experiments at the LHC.  The simplest model  
of this type would be one in which the $\phi$ and $\eta$ Higgs 
self-interactions have similar structures, with a weak $\phi-\eta$ coupling.  
Massiveness of the $\phi$ Higgs states would then imply massiveness of the 
corresponding $\eta$ Higgs states, with only one neutral and two charged 
pseudo Goldstone Higgs states potentially observable at LHC energies.  
In such models, one simultaneously has observable light Higgs states
(the pseudo Goldstone triplet) and ``new physics'' implied by the 
strongly self-coupled $\phi$ and $\eta$ Higgs sectors.  

\bigskip
\centerline{\bf Acknowledgments}
This work was supported in part by the Department of Energy under
Grant \#DE--FG02--90ER40542.   I wish to thank the members of the SLAC 
Theory Group for stimulating comments on a seminar I gave on Ref. [1], 
which led to the writing of this paper.  I also wish to thank C. Bernard, 
B. Mc Kellar, P. Mc Kenzie, C. Quigg, S. Sharpe, and S. B. Treiman  
for helpful conversations or email correspondence. 
\vfill\eject

\centerline{\bf References}
\bigskip
\noindent
\item{[1]}  S. L. Adler, Phys. Rev. D {\bf 59}, 015012 (1999), 
hep-ph/9806518.
\bigskip 
\noindent
\item{[2]}  M. Peskin, private communication.
\bigskip
\noindent
\item{[3]}  S. L. Adler, Erratum for Ref. [1], Phys. Rev. D (in press), and 
available 
by email request to ``adler @ias.edu''.
\item{[4]}  S. L. Glashow and S. Weinberg, Phys. Rev D {\bf 15}, 1958 (1977); 
B. Mc Williams and L.-F. Li, Nucl. Phys. {\bf B179}, 62 (1981); 
O. Shankar, Nucl. Phys. {\bf B206}, 253 (1982).  
\bigskip
\noindent
\item{[5]}  M. K. Gaillard and B. W. Lee, Phys. Rev. D {\bf 10}, 897 (1974), 
equation proceeding Eq.~(2.8).  
\bigskip
\noindent
\item{[6]}  B. Mc Williams and O. Shanker, Phys. Rev. D {\bf 22}, 2853 (1980).
\bigskip
\noindent
\item{[7]}  C. R. Allton, et. al., ``$B$-parameters for $\Delta S=2$ 
Supersymmetric Operators'', hep-lat/9806016.  
\bigskip
\noindent
\item{[8]} Particle Data Group, Eur. Phys. J. C {\bf 3}, 1 (1998).
\bigskip
\noindent
\item{[9]} A. X. El-Khadra, A. S. Kronfeld, P. B. Mackenzie, S. M. Ryan, 
and J. N. Simone, Phys. Rev. D {\bf 58}, 014506 (1998).  
\bigskip
\noindent
\item{[10]}  S. Dawson, ``Introduction to Electroweak Symmetry Breaking'', 
hep-ph/9901280.
\bigskip
\noindent
\item{[11]} R. Gatto, G. Morchio, and F. Strocchi, Phys. Letters {\bf 83B}, 
348 (1979); G. Segr\`e and H. A. Weldon, Ann. Phys. (NY) {\bf 124}, 37 
(1980).  
\vfill\eject
\twelvepoint
\doublespace
\pageno=31
\centerline{\bf Table I.  Six-Higgs doublet model fit to experimental data:} 
\centerline{\bf second case of Sec. V}
\bigskip
\singlespace
$$\vbox {\rm \halign {\hfil #\hfil && \quad \hfil #\hfil \cr
quantity & target value & fitted value\cr
&&\cr
$M_u$ & 0.0033 & 0.0033 \cr
$M_c$ & 1.25 & 1.26 \cr
$M_t$ & 173.8 & 174.0 \cr
&&\cr
$M_d$ & 0.0060 & 0.0064 \cr
$M_s$ & 0.115 & 0.111 \cr
$M_b$ & 4.25 & 4.24 \cr
&&\cr
$M_e$ & 0.00051 & 0.00051 \cr
$M_{\mu}$ & 0.1057 & 0.1057 \cr
$M_{\tau}$ & 1.777  & 1.777 \cr
&&\cr
&&\cr
$s_{12}$ & 0.221 & 0.221 \cr
$s_{13}$ & 0.0059 & 0.0062 \cr
$s_{23}$ & 0.039 & 0.037 \cr
$|\sin \delta_{13}|$ & 0.60 & 0.48 \cr
}}$$                
\vfill\eject
\twelvepoint
\doublespace
\pageno=32
\centerline{\bf Table II.  Six-Higgs doublet model fit to experimental data:} 
\centerline{\bf third case of Sec. V}
\bigskip
\singlespace
$$\vbox {\rm \halign {\hfil #\hfil && \quad \hfil #\hfil \cr
quantity & target value & fitted value\cr
&&\cr
$M_u$ & 0.0033 & 0.0032 \cr
$M_c$ & 1.25 & 1.25 \cr
$M_t$ & 173.8 & 173.9 \cr
&&\cr
$M_d$ & 0.0060 & 0.0065 \cr
$M_s$ & 0.115 & 0.096 \cr
$M_b$ & 4.25 & 4.25 \cr
&&\cr
$M_e$ & 0.00051 & 0.00051 \cr
$M_{\mu}$ & 0.1057 & 0.1057 \cr
$M_{\tau}$ & 1.777  & 1.777 \cr
&&\cr
&&\cr
$s_{12}$ & 0.221 & 0.221 \cr
$s_{13}$ & 0.0059 & 0.0059 \cr
$s_{23}$ & 0.039 & 0.039 \cr
$|\sin \delta_{13}|$ & 0.60 & 0.33 \cr
}}$$                
\bye